\documentclass[pra,aps,superscriptaddress,twocolumn,color]{revtex4-1}

\usepackage{amsmath}
\usepackage{amsfonts}
\usepackage{bm}
\usepackage{graphicx}
\usepackage{array,color}

\begin{document}
\title{Interface properties in three-component Bose-Einstein condensates}

\author{Keisuke Jimbo}
\affiliation{Department of Engineering Science, University of
Electro-Communications, Tokyo 182-8585, Japan}

\author{Hiroki Saito}
\affiliation {Department of Engineering Science, University of
Electro-Communications, Tokyo 182-8585, Japan}

\date{\today}

\begin{abstract}
Interface properties of a three-component Bose-Einstein condensate, in which
component 3 is sandwiched by components 1 and 2 at the interface, are
investigated.
It is shown that component 3 can serve as a surfactant: the net interfacial
tension is reduced by the presence of component 3.
We calculate the interfacial tension as a function of the interaction
coefficients.
The stability of the interface is studied by Bogoliubov analysis.
When the interfacial tension has a spatial gradient, interfacial flow is
induced, which resembles the Marangoni flow.
\end{abstract}

\maketitle

\section{Introduction}
\label{s:intro}

Surfactants lower the interfacial tension between two immiscible fluids.
For example, soap molecules enter the interface between oil and water,
lowering the interfacial tension.
The interfacial tension is decreased because soap molecules have an affinity
for both oil and water, which changes the interface structure.
In this paper, we consider an analogous situation at the interface in
three-component Bose-Einstein condensates (BECs), in which two immiscible
components form an interface and the third component enters that interface.

Multicomponent BECs have been studied extensively both experimentally and
theoretically.
Two-component BECs were realized by using atoms in different hyperfine
states~\cite{Myatt} and different atomic species~\cite{Modugno}.
Miscibility of a two-component BEC depends on the inter- and
intra-component interactions, which can be controlled by the Feshbach
resonance technique~\cite{Papp, Tojo}.
When the two components are immiscible, phase separation occurs and
interfaces are formed~\cite{Hall, Miesner, Mertes, Eto}.
The static and dynamic properties of such interfaces have been investigated
theoretically~\cite{Ho, Pu, Timm, Ao, Tripp, Mazets, Barankov, Schae,
  Takeuchi2, Indekeu, Lee, Thu, Indekeu2}.
The interfacial tension coefficients for an immiscible two-component BEC
were obtained analytically and numerically in Refs.~\cite{Ao, Barankov,
  Schae}.
Dynamical instabilities~\cite{Sasaki, Gautam, Takeuchi, Bezett, Suzuki,
  Kobyakov, Sasaki11, Kadokura, Kobyakov2, Kobyakov3, Sakaguchi, Maity},
such as Rayleigh-Taylor and Kelvin-Helmholtz instabilities, were shown
to deform the interface between two superfluids, as in the case of classical
fluids.
However, all the above studies focus on two-component interfaces;
the interface of a three-component BEC has hitherto not been
explored~\cite{Note}.

\begin{figure}[tb]
\includegraphics[width=5cm]{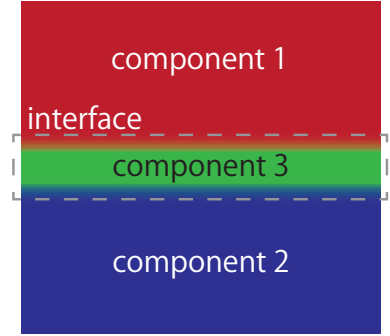}
\caption{
  Schematic of a three-component interface, in which component 3 is
  sandwiched by components 1 and 2.
  The whole region inside the dashed square is regarded as the
  three-component interface.
}
\label{f:schematic}
\end{figure}
Let us suppose components 1 and 2 have segregated to either side of the
interface, and a small amount of component 3 is localized at the interface
(see Fig.~\ref{f:schematic}).
We will also assume that the repulsive interaction between components 1 and
2 is strong, while those between components 1 and 3 and between components 2
and 3 are moderate.
In the absence of component 3, components 1 and 2 are in direct contact at
the interface, in which case the interfacial tension becomes large owing to 
the strong repulsion~\cite{Ao, Barankov, Schae}.
On the other hand, if there is a component 3 lying between components 1 and
2, so that components 1 and 2 have no direct contact, the interfacial
tension is suppressed because of the moderate repulsion between components 1
and 3 and between components 2 and 3.
Thus, the interfacial tension is decreased by the presence of component 3 at
the interface: component 3 can be regarded as a surfactant.

In this paper, we will show numerically and variationally that the above
prediction is true and that the third component can play the role of
surfactant at the interface.
The interfacial tension is calculated for various parameters.
We also perform Bogoliubov analysis to examine the stability of the
three-component interface.
The latter part of this paper investigates the dynamics of the system with a
spatial gradient in interfacial tension.
We will show that a mass current is induced near the interface such that the
gradient in interfacial tension is compensated, similarly to Marangoni
flow~\cite{Marangoni} in classical fluids.

The remainder of the paper is organized as follows.
Section~\ref{s:static} focuses on the static properties of the
three-component interface, including calculations of the interfacial
tension, variational analysis, and Bogoliubov analysis.
Section~\ref{s:dynamic} numerically investigates the dynamics of Marangoni
flow.
Section~\ref{s:conc} presents the conclusions of this study.

\section{Static properties}
\label{s:static}

\subsection{Interfacial tension}
\label{s:interface}

We consider a three-component BEC of dilute gases in a uniform space at zero
temperature.
In the mean-field approximation, the energy of the system is given by
\begin{eqnarray} \label{E}
  E & = & \int d\bm{r} \Biggl[ \sum_{j=1}^3 \left( -\psi_j^*
    \frac{\hbar^2}{2 m_j}   \nabla^2 \psi_j + \frac{g_{jj}}{2} |\psi_j|^4
    \right) \nonumber \\
& & + \sum_{j < j'} g_{jj'} |\psi_j|^2 |\psi_{j'}|^2 \Biggr],
\end{eqnarray}
where $\psi_j(\bm{r})$ is the macroscopic wave function and $m_j$ is the
atomic mass of component $j$ ($j = 1$, 2, and 3).
The interaction coefficients in Eq.~(\ref{E}) are defined as $g_{jj'} = 2\pi
\hbar^2 a_{jj'} / m_{jj'}$, where $a_{jj'}$ is the $s$-wave scattering
length and $m_{jj'}$ is the reduced mass between components $j$ and $j'$.
We treat the problem in the grand canonical ensemble, and the grand
potential,
\begin{equation} \label{Omega}
\Omega = E - \sum_{j=1}^3 \mu_j N_j,
\end{equation}
is minimized in the equilibrium state, where $\mu_j$ is the chemical
potential and
\begin{equation}
N_j = \int d\bm{r} |\psi_j(\bm{r})|^2
\end{equation}
is the number of atoms in component $j$.
The macroscopic wave functions $\psi_j(\bm{r})$ in the equilibrium state
thus obey $\delta \Omega / \delta\psi^*(\bm{r}) = 0$, which gives the
coupled Gross-Pitaevskii (GP) equations,
\begin{equation} \label{GP}
-\frac{\hbar^2}{2m_j} \nabla^2 \psi_j + \sum_{j' = 1}^3 g_{jj'}
|\psi_{j'}|^2 \psi_j = \mu_j \psi_j.
\end{equation}

The miscibility of multicomponent BECs is determined by the interaction
coefficients $g_{jj'}$.
The uniformly mixed state of components $j$ and $j'$ is unstable against
phase separation, when the inequality
\begin{equation} \label{im}
g_{jj'}^2 > g_{jj} g_{j'j'}
\end{equation}
is satisfied~\cite{Pethick}.
This immiscibility condition can also be extended to the three-component
BEC~\cite{Roberts}: the three components are immiscible with each other,
when all pairs of components satisfy Eq.~(\ref{im}), that is,
\begin{equation} \label{im3}
  g_{12}^2 > g_{11} g_{22} \qquad
  g_{13}^2 > g_{11} g_{33}, \qquad
  g_{23}^2 > g_{22} g_{33}.
\end{equation}
In this paper, we assume that the interaction coefficients satisfy this
immiscibility condition for a three-component BEC.

We consider a situation in which components 1 and 2 have uniform densities
$n_{1\infty}$ and $n_{2\infty}$ at $z = -\infty$ and $z = \infty$,
respectively,
and the interface between them is located at $z = 0$, around which component
3 is localized.
The system is assumed to be uniform in the $x$ and $y$ directions and the GP
equation (\ref{GP}) reduces to
\begin{equation} \label{GPz}
-\frac{\hbar^2}{2m_j} \psi_j''(z) + \sum_{j' = 1}^3 g_{jj'}
|\psi_{j'}(z)|^2 \psi_j(z) = \mu_j \psi_j(z)
\end{equation}
with $\mu_1 = g_{11} n_{1\infty}$ and $\mu_2 = g_{22} n_{2\infty}$, where
the boundary conditions are $\psi_1(z = -\infty) = \sqrt{n_{1\infty}}$ and
$\psi_2(z = \infty) = \sqrt{n_{2\infty}}$.
The pressures on either side of the interface must be balanced with each
other as
\begin{equation}
\frac{g_{11} n_{1\infty}^2}{2} = \frac{g_{22} n_{2\infty}^2}{2} \equiv P.
\end{equation}
Multiplying Eq.~(\ref{GPz}) by $\psi_j'(z)$ and integrating the sum of the
three equations with respect to $z$, we obtain
\begin{equation} \label{integrate}
\sum_{j=1}^3 \left[ -\frac{\hbar^2}{2m_j} (\psi_j')^2 + \frac{g_{jj}}{2}
  \psi_j^4 - \mu_j \psi_j^2 \right] + \sum_{j>j'} g_{jj} \psi_j \psi_{j'} +
C = 0,
\end{equation}
where $C$ is an integration constant and $\psi_j(z)$ is taken to be real
without loss of generality.
In the limit $z \rightarrow \pm \infty$, the first summation in
Eq.~(\ref{integrate}) becomes $-P$ and the second summation vanishes;
therefore, the constant $C$ must be $P$.
Substitution of Eq.~(\ref{integrate}) into Eq.~(\ref{Omega}) gives
\begin{equation} \label{Omega2}
\Omega = -PV + 2 \int d\bm{r} \sum_{j=1}^3 \frac{\hbar^2}{2m_j} (\psi_j')^2,
\end{equation}
where $V$ is the volume of the system.
Since $\psi_j'(z)$ is nonzero only near the interface and vanishes elsewhere,
the second term on the right-hand side of Eq.~(\ref{Omega2}) arises only
from the interface region.
Thus, the interfacial tension $\sigma$ can be expressed as
\begin{equation} \label{sigma}
\sigma = 2 \int_{-\infty}^\infty dz \sum_{j=1}^3 \frac{\hbar^2}{2m_j}
(\psi_j')^2.
\end{equation}
This expression is similar to the one derived for two-component
systems~\cite{Schae}.

We numerically solve the GP equation~(\ref{GPz}) to obtain the equilibrium
state $\psi_j(z)$, and using this, we calculate the interfacial tension of
the three-component interface by Eq.~(\ref{sigma}).
In order to reduce the number of parameters, we assume $m_1 = m_2 = m_3
\equiv m$ and $g_{11} = g_{22} = g_{33} \equiv g$ in the following
calculations.
We also assume $\mu_1 = \mu_2$, and hence $n_{1\infty} = n_{2\infty} \equiv
n_\infty$.
We normalize length, time, energy, and density by $\xi = \hbar /
(2mgn_\infty)^{1/2}$, $\hbar / (gn_\infty)$, $gn_\infty$, and $n_\infty$,
respectively.
In this unit, $g_{11} = g_{22} = g_{33} = 1$ and $\mu_1 = \mu_2 = 1$.
The variable parameters are thus $g_{12}$, $g_{13}$, $g_{23}$, and $\mu_3$.
To obtain the solution for Eq.~(\ref{GPz}), we solve the imaginary-time GP
equation as
\begin{equation} \label{imagtime}
  \frac{\partial \psi_j}{\partial \tau} = \frac{1}{2} \psi_j''
  - \sum_{j' = 1}^3 g_{jj'} |\psi_{j'}|^2 \psi_j + \mu_j \psi_j,
\end{equation}
using the pseudospectral method~\cite{recipe}.
The initial state of the imaginary-time evolution is taken to be an
appropriate state, e.g., $\psi_1(z) = \theta(-z)$, $\psi_2(z) = \theta(z)$,
and $\psi_3(z) = \exp(-z^2)$, so that it converges to the desired interface
state for large imaginary times, where $\theta(z)$ is the Heaviside step
function.
We take a large enough space so that the boundary does not affect the
interface region.

\begin{figure}[tb]
\includegraphics[width=8cm]{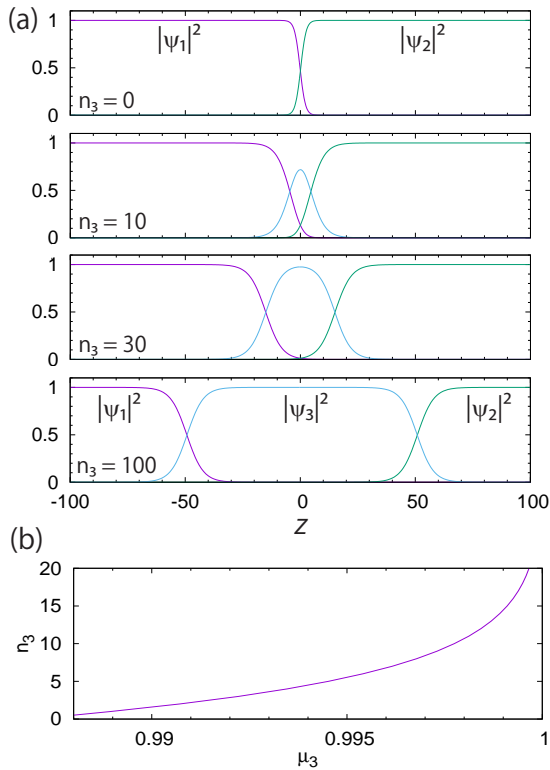}
\caption{
Flat-interface equilibrium states.
(a) Density profile of each component near the interface for $n_3 = 0$, 10,
30, and 100.
(b) Areal density $n_3$ as a function of the chemical potential $\mu_3$ for
component 3.
The parameters are $g_{12} = 1.1$ and $g_{13} = g_{23} = 1.01$.
}
\label{f:shape}
\end{figure}
Figure~\ref{f:shape}(a) shows the density profile $|\psi_j(z)|^2$ around the
interface for different values of $\mu_3$.
The intercomponent interaction coefficients are taken to be $g_{12} = 1.1$
and $g_{13} = g_{23} = 1.01$, which satisfy the immiscibility condition in
Eq.~(\ref{im3}).
For $\mu_3 = 0$, component 3 vanishes, and a two-component (1 and 2)
interface is realized, as shown in the uppermost panel in
Fig.~\ref{f:shape}(a).
The number of atoms $n_3$ in component 3 per unit area of the interface is
given by
\begin{equation}
n_3 = \int |\psi_3(z)|^2 dz,
\end{equation}
which is a monotonically increasing function of $\mu_3$, as shown in
Fig.~\ref{f:shape}(b).
(Note that the numerical values of $n_3$ are normalized by $n_{\infty}
\xi$ in the present unit of normalization.)
As $n_3$ increases, component 3 localized near the interface pushes away
components 1 and 2, and components 1 and 2 are detached from each other
($n_3 = 30$ in Fig.~\ref{f:shape}(a)).
When $n_3$ increases further ($n_3 = 100$ in Fig.~\ref{f:shape}(a)),
component 3 becomes a plateau, which separates the two distinct (1-3 and
2-3) interfaces.
In this paper, such a structure, as a whole, is also regarded as a single
three-component interface.

\begin{figure}[tb]
\includegraphics[width=8cm]{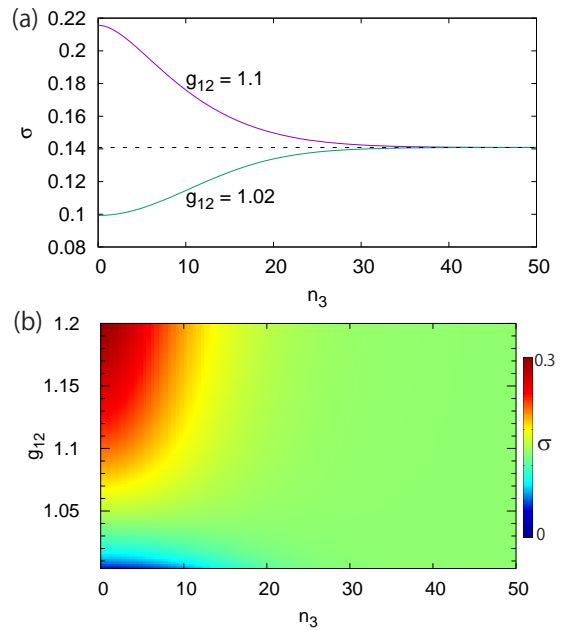}
\caption{
  Interfacial tension $\sigma$ for $g_{13} = g_{23} = 1.01$.
  (a) $n_3$ dependence for $g_{12} = 1.1$ and $g_{12} = 1.02$.
  The horizontal dashed line represents $\sigma_{\rm binary}(g_{13})
  + \sigma_{\rm binary}(g_{23}) \simeq 0.14$.
  (b) Dependence of $\sigma$ on $g_{12}$ and $n_3$.
}
\label{f:sigma}
\end{figure}
We calculate the interfacial tension $\sigma$ using Eq.~(\ref{sigma}) for
the equilibrium state.
Figure~\ref{f:sigma}(a) shows $\sigma$ as a function of $n_3$.
The case of $g_{12} = 1.1$ corresponds to the condition in
Fig.~\ref{f:shape}.
In this case, the interfacial tension $\sigma$ decreases with increasing
$n_3$.
Since the interfacial tension is decreased by the presence of component 3,
we can say that component 3 plays the role of surfactant.
By contrast, for $g_{12} = 1.02$, $\sigma$ increases with $n_3$.
Figure~\ref{f:sigma}(b) shows the dependence of $\sigma$ on $g_{12}$ and
$n_3$.
The critical value of $g_{12}$ for the surfactant behavior is $g_{12} \simeq
1.04$, above which $\sigma$ is a decreasing function of $n_3$.

We denote the interfacial tension of the two-component interface by
$\sigma_{\rm binary}(g_{\rm inter})$, as a function of the intercomponent
interaction coefficient $g_{\rm inter}$ (the intracomponent interaction
coefficients being normalized to unity).
When $g_{\rm inter}$ satisfies $g_{\rm inter} - 1 \ll 1$, an approximate
expression for the interfacial tension $\sigma_{\rm binary}$ of a
two-component BEC can be derived as~\cite{Ao, Barankov, Schae}
(in the present unit of normalization)
\begin{equation} \label{approx}
  \sigma_{\rm binary}(g_{\rm inter}) \simeq
  \sqrt{\frac{g_{\rm inter} - 1}{2}}.
\end{equation}
In Fig.~\ref{f:sigma}(a), the interfacial tension at $n_3 = 0$ is $\sigma
\simeq 0.216$ for $g_{12} = 1.1$ and $\sigma \simeq 0.099$ for $g_{12} =
1.02$, namely, $\sigma_{\rm binary}(1.1) \simeq 0.216$ and
$\sigma_{\rm binary}(1.02) \simeq 0.099$.
These values agree well with the approximation in Eq.~(\ref{approx}):
$\sqrt{(1.1 - 1) / 2} \simeq 0.224$ and
$\sqrt{(1.02 - 1) / 2} \simeq 0.1$.
In Fig.~\ref{f:sigma}(a), the interfacial tension converges to $\sigma
\simeq 0.141$ for large $n_3$.
This value of the interfacial tension should be $\sigma_{\rm binary}(g_{13})
+ \sigma_{\rm binary}(g_{23}) = 2\sigma_{\rm binary}(1.01)$, since the
three-component interface consists of two separate interfaces (1-3 and 2-3
interfaces) for large $n_3$, as shown in the bottom panel in
Fig.~\ref{f:shape}(a).
Using Eq.~(\ref{approx}) with $g_{13} = g_{23} = 1.01$, we get
$\sigma_{\rm binary}(g_{13}) + \sigma_{\rm binary}(g_{23}) \simeq 2
\sqrt{(1.01 - 1) / 2} \simeq 0.141$, which is in good agreement with the
value obtained above.

The surfactant behavior of component 3 is understood as follows.
When component 3 is absent, i.e., when components 1 and 2 are in direct
contact, the interfacial tension $\sigma = \sigma_{\rm inter}({g_{12}})$ is
large, since $g_{12}$ is large.
This is because the one component cannot penetrate into the other for large
$g_{12}$, and the densities have to change abruptly at the interface, which
results in a large interfacial tension according to Eq.~(\ref{sigma}).
When component 3 is inserted into the interface, component 3 mediates
between components 1 and 2.
Since $g_{13}$ and $g_{23}$ are not so large, the density variation in each
component is moderate across this 1-3-2 interface.
In fact, the slopes of $|\psi_1|^2$ and $|\psi_2|^2$ at the interface in
Fig.~\ref{f:shape}(a) for $n_3 \geq 10$ are more gradual than those for $n_3
= 0$.
Thus, the interfacial tension is decreased by component 3.

\begin{figure}[tb]
\includegraphics[width=9cm]{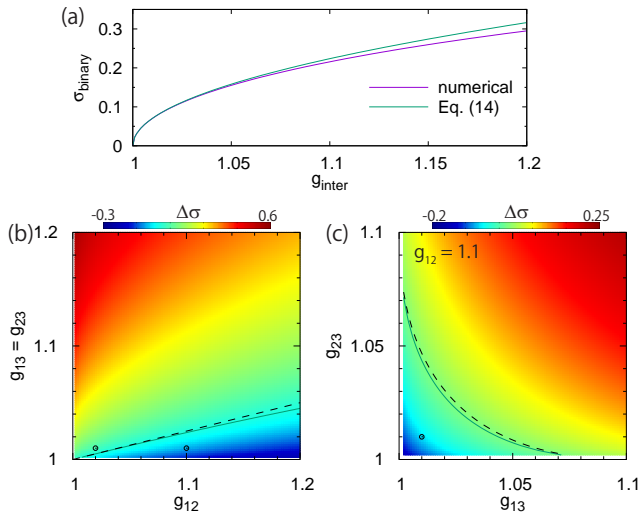}
\caption{
  (a) Interfacial tension $\sigma_{\rm binary}$ for a two-component BEC as a
  function of the intercomponent interaction $g_{\rm inter}$.
  The numerical result is compared with Eq.~(\ref{approx}).
  (b) and (c) Dependence of $\Delta\sigma$ in Eq.~(\ref{delta}) on the
  intercomponent interactions $g_{12}$, $g_{23}$, and $g_{13}$, where the
  values of $\sigma_{\rm binary}$ in Eq.~(\ref{delta}) are taken from
  the numerical result in (a).
  In (c), $g_{12}$ is fixed to 1.1. 
  The solid lines represent the contour of $\Delta\sigma = 0$, and the
  dashed lines are Eq.~(\ref{delta2}).
  The points correspond to the parameters used in Fig.~\ref{f:sigma}(a).
}
\label{f:delta}
\end{figure}
As shown in Fig.~\ref{f:shape}(a), $\sigma = \sigma_{\rm binary}(g_{12})$
for $n_3 = 0$ and $\sigma$ converges to $\sigma_{\rm binary}(g_{13}) +
\sigma_{\rm binary}(g_{23})$ for large $n_3$. 
Therefore, their difference,
\begin{equation} \label{delta}
  \Delta\sigma \equiv \sigma_{\rm binary}(g_{13})
  + \sigma_{\rm binary}(g_{23}) - \sigma_{\rm binary}(g_{12}),
\end{equation}
characterizes the surfactant behavior of the system;
component 3 serves as a surfactant when $\Delta\sigma < 0$.
Equation~(\ref{delta}) only includes the interfacial tension of a
two-component BEC.
We numerically obtain $\sigma_{\rm binary}(g_{\rm inter})$ as a function of
$g_{\rm inter}$ using the two-component version of Eqs.~(\ref{sigma}) and
(\ref{imagtime}), which is shown in Fig.~\ref{f:delta}(a).
The numerical result agrees well with the approximate expression in
Eq.~(\ref{approx}) for $g_{\rm inter}$ close to unity. 
Using the numerically obtained $\sigma_{\rm binary}(g_{\rm inter})$, we
calculate $\Delta\sigma$ in Eq.~(\ref{delta}), which is plotted in
Figs.~\ref{f:delta}(b) and \ref{f:delta}(c).
The solid lines in Figs.~\ref{f:delta}(b) and \ref{f:delta}(c) represent the
contours of $\Delta\sigma = 0$, and the surfactant regions of $\Delta\sigma
< 0$ lie below these lines.
Using the approximate expression in Eq.~(\ref{approx}), $\Delta \sigma = 0$
reads
\begin{equation} \label{delta2}
\sqrt{g_{13} - 1} + \sqrt{g_{23} - 1} - \sqrt{g_{12} - 1} = 0,
\end{equation}
which is represented by the dashed lines in Figs.~\ref{f:delta}(b) and
\ref{f:delta}(c).
These are in good agreement with the solid lines.

\subsection{Variational analysis}
\label{s:var}

We perform variational analysis for better understanding of the numerical
results in Sec.~\ref{s:interface}.
We use the following form of the variational wave functions:
\begin{subequations} \label{var}
\begin{eqnarray}
  \psi_1^{\rm (var)}(z) & = & \sqrt{\frac{1}{2} \left( 1 - \tanh
    \frac{z + z_0}{\alpha} \right)}, \\
  \psi_2^{\rm (var)}(z) & = & \sqrt{\frac{1}{2} \left( 1 + \tanh
    \frac{z - z_0}{\alpha} \right)}, \\
  \psi_3^{\rm (var)}(z) & = & \sqrt{\frac{1}{2} \left(
    \tanh \frac{z + w / 2}{\alpha} - \tanh \frac{z - w / 2}{\alpha}
    \right)}, \nonumber \\
  \label{var3}
\end{eqnarray}
\end{subequations}
where $z_0$ and $\alpha$ are variational parameters characterizing the
position and width of the interface.
At infinity $z \rightarrow \pm\infty$, these variational functions have the
same limiting behaviors as the wave functions in Sec.~\ref{s:interface},
i.e., $\psi_1^{\rm (var)} \rightarrow 1$ and $\psi_2^{\rm (var)}
\rightarrow 0$ for $z \rightarrow -\infty$, and $\psi_1^{\rm (var)}
\rightarrow 0$ and $\psi_2^{\rm (var)} \rightarrow 1$ for $z \rightarrow
\infty$.
In Eq.~(\ref{var3}), the parameter $w$ determines the number of atoms in
component 3, because
\begin{equation}
\int_{-\infty}^\infty dz |\psi_3^{\rm (var)}(z)|^2 = w.
\end{equation}
If $z_0 = w$, Eq.~(\ref{var}) gives the uniform total density,
$|\psi_1^{\rm (var)}(z)|^2 + |\psi_2^{\rm (var)}(z)|^2
+ |\psi_3^{\rm (var)}(z)|^2 = 1$.
Substituting the variational wave functions in Eq.~(\ref{var}) into the
grand potential in Eq.~(\ref{Omega}), and subtracting the divergent part, we
obtain the variational grand potential $\Omega_{\rm var}$ (see
Appendix~\ref{app} for derivation).
We minimize $\Omega_{\rm var}$ with respect to $z_0$ and $\alpha$.
Using these optimal values of $z_0$ and $\alpha$, we calculate the
interfacial tension via
\begin{eqnarray}
\sigma & = & 2 \int_{-\infty}^\infty dz \sum_{j=1}^3 \frac{1}{2}
\left( \frac{d\psi_j^{\rm (var)}}{dz} \right)^2 \nonumber \\
& = & \frac{1}{2\alpha} + \frac{1}{2\alpha^2} \left[ \alpha \coth
  \frac{w}{\alpha} - w \left( \sinh \frac{w}{\alpha} \right)^{-2} \right].
\end{eqnarray}

\begin{figure}[tb]
\includegraphics[width=8cm]{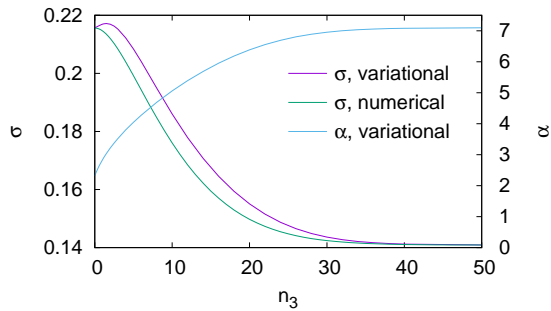}
\caption{
  Comparison between the interfacial tension $\sigma$ obtained by the
  variational method (``$\sigma$, variational'') and that obtained by
  solving the GP equation (``$\sigma$, numerical'', the same data as in
  Fig.~\ref{f:sigma}(a)) for $g_{12} = 1.1$ and $g_{13} = g_{23} = 1.01$.
  The variational parameter $\alpha$ minimizing the variational energy
  is also shown (``$\alpha$, variational'').
}
\label{f:var}
\end{figure}
Figure~\ref{f:var} compares the interfacial tension obtained by the
variational method with that obtained in Fig.~\ref{f:sigma}(a) numerically.
The variationally obtained $\sigma$ agrees well with the numerically
obtained $\sigma$ for $n_3 = 0$ and $n_3 \gtrsim 40$.
In the intermediate region of $n_3$, the former deviates slightly from the
latter, which is due to the simple form of the variational wave function in
Eq.~(\ref{var}).
The optimal variational parameter $\alpha$ that corresponds to the interface
width is also shown in Fig.~\ref{f:var}.
The value of $\alpha$ increases with $n_3$, and therefore, our variational
wave function captures the basic physics of the surfactant behavior:
the spatial variation in the densities at the interface is made gradual by
component 3.

\subsection{Thermodynamic relations}

According to the definition of interfacial tension $\sigma$, the work
required for increasing the interface area $A$ by $dA$ is given by $\sigma 
dA$.
Thus, at zero temperature and in a fixed volume, the energy differential is
given by $dE = \sum_j \mu_j dN_j + \sigma dA$, and therefore the
differential of the grand potential has the form,
\begin{equation} \label{dOmega}
d\Omega = -\sum_{j=1}^3 N_j d\mu_j + \sigma dA.
\end{equation}
This differential relation leads to
\begin{equation} \label{dOdmu}
\frac{\partial \Omega}{\partial \mu_j} = -N_j.
\end{equation}
In general, the grand potential $\Omega$ can be divided into a volume part
and an interface part $\sigma A$, as shown in Eq.~(\ref{Omega2}) for a flat
interface, where the interface part corresponds to the second term on the
right-hand side of Eq.~(\ref{Omega2}).
Since component 3 is localized at the interface, $\mu_3$ is included only in
the interface part $\sigma A$ of the grand potential, and therefore,
\begin{equation} \label{dOdmu2}
  \frac{\partial \Omega}{\partial \mu_3} =
  A \frac{\partial \sigma}{\partial \mu_3}.
\end{equation}
From Eqs.~(\ref{dOdmu}) and (\ref{dOdmu2}), we obtain
\begin{equation} \label{dsdmu}
  \frac{\partial \sigma}{\partial \mu_3} = -\frac{N_3}{A}.
\end{equation}
We have numerically confirmed that this relation holds for
Fig.~\ref{f:shape}(b), i.e., $d \sigma / d \mu_3 = n_3$.

We assume that the volume part of the system is sufficiently large, and
$N_1, N_2 \rightarrow \infty$ with $\mu_1$ and $\mu_2$ being fixed.
In this case, $\sigma$ is a function of only $\mu_3$ or $N_3$, as in the
case of Fig.~\ref{f:shape}.
It follows from Eq.~(\ref{dsdmu}) that
\begin{equation}
  \frac{d\sigma}{d N_3} = -\frac{N_3}{A} \frac{d \mu_3}{d N_3}.
\end{equation}
Integrating this equation with respect to $N_3$, we have
\begin{equation}
  \sigma = -\frac{1}{A} \int N_3 \frac{d\mu_3}{dN_3} dN_3 + c,
\end{equation}
where we determine the integration constant $c$ appropriately.
When $N_3 = 0$, $\sigma$ should be the two-component interfacial tension
$\sigma_{\rm binary}(g_{12})$, and hence,
\begin{equation} \label{sigma1}
  \sigma = -\frac{1}{A} \int_0^{N_3} N_3 \frac{d\mu_3}{dN_3}
  dN_3 + \sigma_{\rm binary}(g_{12}).
\end{equation}
In the limit of $N_3 \rightarrow \infty$, the interfacial tension should
approach $\sigma \rightarrow \sigma_{\rm binary}(g_{13})
+ \sigma_{\rm binary}(g_{23})$, giving
\begin{equation} \label{sigma2}
  \sigma = \frac{1}{A} \int_{N_3}^\infty N_3
  \frac{d\mu_3}{dN_3} dN_3 + \sigma_{\rm binary}(g_{13})
  + \sigma_{\rm binary}(g_{23}).
\end{equation}
Using Eqs.~(\ref{sigma1}) and (\ref{sigma2}), $\Delta\sigma$ defined in
Eq.~(\ref{delta}) can be written as
\begin{equation} \label{ds}
  \Delta\sigma = -\frac{1}{A} \int_0^\infty N_3 \frac{d\mu_3}{dN_3} dN_3.
\end{equation}
A similar relation is derived for the wetting problem in Ref.~\cite{Landau}.
Interestingly, $\Delta\sigma$ in Eq.~(\ref{ds}) is expressed in terms of
only $\mu_3$ and $N_3$.
We have numerically confirmed that the relation (\ref{ds}) is satisfied for
the flat interface in Sec.~\ref{s:interface}.

\subsection{Bogoliubov analysis}

We perform Bogoliubov analysis to study the dynamical stability of the
flat interface discussed in Sec.~\ref{s:interface}.
We consider a small deviation $\delta\psi_j(\bm{r}, t)$ from the
stationary state $f_j(z)$ with a flat interface located around the $z = 0$
plane, and therefore, the wave function can be broken down into
\begin{equation} \label{bogopsi}
\psi_j(\bm{r}, t) = e^{-i \mu_j t / \hbar} \left[ f_j(z) +
  \delta\psi_j(\bm{r}, t) \right],
\end{equation}
where $f_j(z)$ is the solution for Eq.~(\ref{GPz}) obtained by the
imaginary-time evolution of Eq.~(\ref{imagtime}).
Substitution of Eq.~(\ref{bogopsi}) into the time-dependent GP equation,
\begin{equation}
i \frac{\partial\psi_j}{\partial t} = 
-\frac{1}{2} \nabla^2 \psi_j + \sum_{j' = 1}^3 g_{jj'} |\psi_{j'}|^2 \psi_j,
\end{equation}
gives
\begin{eqnarray}
  i \frac{\partial\delta\psi_j}{\partial t} & = & -\frac{1}{2} \nabla^2
  \delta\psi_j - \mu_j \delta\psi_j + \sum_{j' = 1}^3 g_{jj'} \bigl(
  |f_{j'}|^2 \delta\psi_j 
\nonumber \\
& & + f_{j'}^* f_j \delta\psi_{j'} + f_{j'} f_j \delta\psi_{j'}^* \bigr).
\label{bogo1}
\end{eqnarray}
The small deviation $\delta\psi_j(\bm{r}, t)$ can be decomposed into
Fourier components, and we focus on a plane wave with wave number
$\bm{k}_\perp$ along the interface:
\begin{equation} \label{dpsi}
  \delta\psi_j(\bm{r}, t) =
  u_j(z) e^{i \bm{k}_\perp \cdot \bm{r}_\perp - i \omega t}
  + v_j^*(z) e^{-i \bm{k}_\perp \cdot \bm{r}_\perp + i \omega^* t},
\end{equation}
where the subscript $\perp$ indicates that the vector is in the $x$-$y$
plane.
Substituting Eq.~(\ref{dpsi}) into Eq.~(\ref{bogo1}), we obtain the
following Bogoliubov-de Gennes equations:
\begin{subequations} \label{BdG}
\begin{eqnarray}
  \omega u_j & = & -\frac{1}{2} u_j'' + \left( \frac{k_\perp^2}{2} - \mu_j
  \right) u_j + \sum_{j'=1}^3 g_{jj'} \bigl( |f_{j'}|^2 u_j \nonumber \\
  & & + f_j f_{j'}^* u_{j'} + f_j f_{j'} v_{j'} \bigr), \\
  -\omega v_j & = & -\frac{1}{2} v_j'' + \left( \frac{k_\perp^2}{2} - \mu_j
  \right) v_j + \sum_{j'=1}^3 g_{jj'} \bigl( |f_{j'}|^2 v_j \nonumber \\
  & & + f_j f_{j'}^* v_{j'} + f_j f_{j'} u_{j'} \bigr),
\end{eqnarray}
\end{subequations}
which are six simultaneous differential equations.

We spatially discretize $u_j(z)$ and $v_j(z)$ in Eq.~(\ref{BdG}) and
approximate the second derivative by $u_j'' \simeq [u_j(z + \delta z) - 2
u_j(z) + u_j(z + \delta z)] / \delta z^2$, which gives the matrix form of
the eigenvalue equation.
We numerically diagonalize the eigenvalue equation to obtain the
eigenfrequency $\omega$.
If $\omega$ is real for any $k_\perp$, the stationary flat interface state
is dynamically stable.
If the imaginary part of $\omega$ is nonzero, the corresponding mode grows
in time exponentially, and the system is dynamically unstable.

\begin{figure}[tb]
\includegraphics[width=8cm]{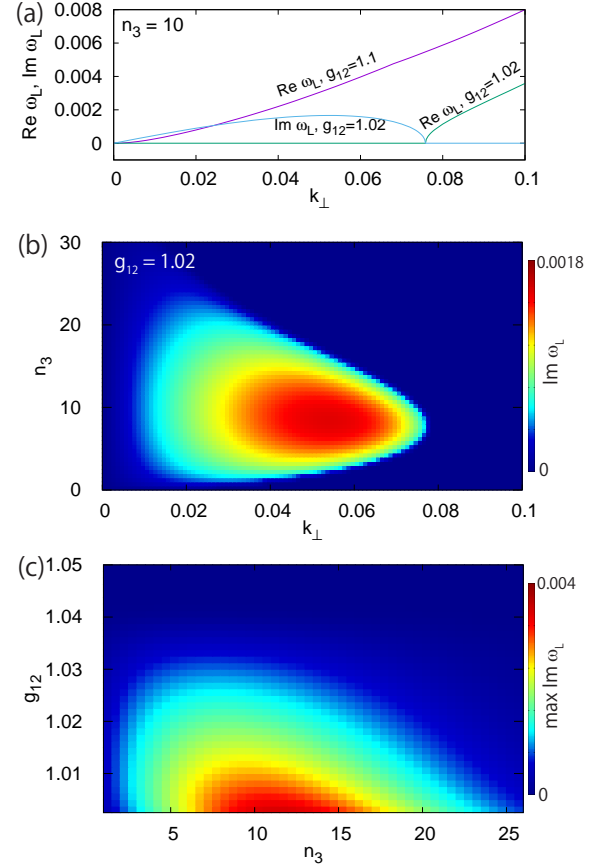}
\caption{
  Lowest excitation frequency $\omega_L$ obtained by numerically
  diagonalizing Eq.~(\ref{BdG}).
  (a) Real and imaginary parts of $\omega_L$ as functions of $k_\perp$ for
  $g_{12} = 1.1$, $g_{12} = 1.02$, and $n_3 = 10$.
  For $g_{12} = 1.1$, ${\rm Im} \omega = 0$ for all the excitations.
  (b) Imaginary part of $\omega_L$ as a function of $k_\perp$ and $n_3$ for
  $g_{12} = 1.02$.
  (c) Largest imaginary part max ${\rm Im} \omega_L$ with respect to
  $k_\perp$ as a function of $n_3$ and $g_{12}$.
  In (a)-(c), $g_{13}$ and $g_{23}$ are fixed to 1.01.
}
\label{f:bogo}
\end{figure}
Figure~\ref{f:bogo} shows the lowest excitation frequency $\omega_L$ that
has the smallest real part among the frequencies $\omega$.
For the parameters investigated in Fig.~\ref{f:bogo}, only $\omega_L$
becomes complex and other $\omega$ are found to be real.
In Fig.~\ref{f:bogo}(a), we first examine the interaction parameters used in
Fig.~\ref{f:sigma}(a) with $n_3 = 10$ being fixed.
For $g_{12} = 1.1$, which corresponds to the surfactant behavior in
Fig.~\ref{f:sigma}(a), ${\rm Re} \omega_L$ increases with $k_\perp$ while
${\rm Im} \omega_L = 0$, and the flat interface is dynamically stable.
We confirmed that this stability is maintained for any $n_3$, and the
interface with surfactant behavior considered in Fig.~\ref{f:sigma}(a) is
always stable.
For $g_{12} = 1.02$, on the other hand, $\omega_L$ is purely imaginary for
$0 < k_\perp \lesssim 0.076$.
The imaginary part ${\rm Im} \omega_L$ is the largest for $k_\perp \simeq
0.05$, and the interface is the most unstable in this wavelength.

Figure~\ref{f:bogo}(b) shows ${\rm Im} \omega_L$ as a function of $k_\perp$
and $n_3$ for $g_{12} = 1.02$, and Fig.~\ref{f:bogo}(c) shows the largest
imaginary part max ${\rm Im} \omega_L(k_\perp)$ among all $k_\perp$ as a
function of $n_3$ and $g_{12}$.
Figures~\ref{f:bogo}(b) and \ref{f:bogo}(c) show that the interface
stabilizes for $n_3 = 0$ and $n_3 \gtrsim 30$.
This is deduced from the fact that an isolated two-component interface is
always stable;
for $n_3 = 0$, the interface reduces to the 1-2 interface, and for large
$n_3$, the 1-3 and 2-3 interfaces separate and can be regarded two distinct
two-component interfaces.
It follows from this result that instability arises from the interaction
among the three components, including the interaction between components 1
and 2 across the layer of component 3.
After dynamical instability sets in, droplets of component 3 are formed,
as seen in capillary instability~\cite{Sasaki11}.

\section{Dynamic Properties: Marangoni flow}
\label{s:dynamic}

When there is a spatial gradient in the interfacial tension $\sigma$ along
the interface, the interface with larger $\sigma$ pulls the interface with
smaller $\sigma$, which results in a flow along the interface.
Such an interface flow driven by the interfacial tension gradient is
referred to as a Marangoni flow~\cite{Marangoni}.
The Marangoni effect can be observed in daily life: for example, tears of
wine~\cite{Thomson} and stabilization of soap films.
The Marangoni effect is also important in the formation of B\'enard cells in
convection.
In these examples, the gradient in the surface or interfacial tension is
caused by the gradients in solute concentration and in temperature.

\begin{figure}[tb]
\includegraphics[width=8cm]{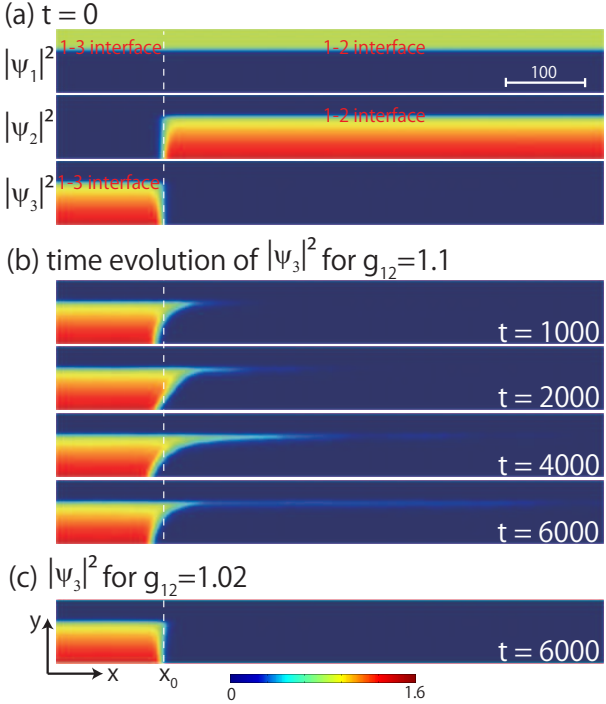}
\caption{
  Dynamics of interface flow due to a gradient in interfacial tension:
  Marangoni flow. 
  (a) Initial density distribution of the three components, which is the
  ground state for $g_{12} = g_{13} = g_{23} = 1.01$.
  (b) Time evolution after $g_{12}$ is changed from 1.01 to 1.1 at $t = 0$.
  (c) Snapshot at $t = 6000$, where $g_{12}$ is changed from 1.01 to 1.02 at
  $t = 0$.
  The vertical dashed line indicates the initial position of the 2-3
  interface $x_0$.
  See supplemental material for videos of the dynamics in (b) and
  (c)~\cite{movies}.
}
\label{f:marangoni}
\end{figure}
Here we numerically demonstrate the Marangoni-like flow in the
three-component BEC in a two-dimensional system.
To realize the interfacial tension gradient along the interface, we consider
the situation shown in Fig.~\ref{f:marangoni}(a).
The 1-2 and 1-3 interfaces are straight along the $x$ direction, and
components 2 and 3 are located in the regions of $x \gtrsim x_0$ and $x
\lesssim x_0$, respectively, where $x_0$ is the position of the 2-3
interface.
The interaction coefficients are initially taken to be $g_{12} = g_{13}$,
and the interfacial tensions of the 1-2 and 1-3 interfaces are the same in
the initial state.
At $t = 0$, we suddenly increase $g_{12}$, which increases the interfacial
tension of the 1-2 interface.
Consequently, the horizontal interface acquires a gradient along the $x$
direction, i.e., the interfacial tension at $x \gtrsim x_0$ exceeds that at
$x \lesssim x_0$, and we presume that the 1-2 interface entrains the 1-3
interface in the $+x$ direction.

We perform a numerical simulation of the time-dependent GP equation:
\begin{equation} \label{tGP}
i \frac{\partial\psi_j}{\partial t} = -\frac{1}{2} \nabla^2 \psi_j + V_j
\psi_j + \sum_{j'=1}^3 g_{jj'} |\psi_{j'}|^2 \psi_j.
\end{equation}
The initial state is the ground state for $g_{12} = g_{13} = g_{23} = 1.01$,
as shown in Fig.~\ref{f:marangoni}(a), which is obtained by the
imaginary-time evolution of Eq.~(\ref{tGP}).
The number of atoms $N_j$ in each component is chosen such that the density
becomes $|\psi_j|^2 \simeq 1$ near the horizontal interface.
At $t = 0$, $g_{12}$ is increased abruptly, and the real-time evolution is
calculated.
To maintain a straight interface through the imaginary- and real-time
evolutions, we apply a potential $V_2 = V_3 = 0.01 y$ ($V_1 = 0$), which
exerts a weak force on components 2 and 3 in the $-y$ direction.
Because the size of the whole system is sufficiently large, the boundary
does not affect the dynamics shown.

Figure~\ref{f:marangoni}(b) shows the time evolution of the density
distribution of component 3, where $g_{12}$ is changed from 1.01 to 1.1 at
$t = 0$.
As a result of this change, the interfacial tension at the 1-2 interface
becomes $\sigma = \sigma_{\rm binary}(1.1) \simeq 0.216$, while $\sigma =
\sigma_{\rm binary}(1.01) \simeq 0.071$ at the 1-3 interface.
As expected, the 1-2 interface pulls the 1-3 interface into the $+x$
direction, and component 3 flows into the $+x$ direction near the interface.
This interface flow leads to a three-component (1-3-2) interface in the
region $x > x_0$.
The interfacial tension of this 1-3-2 interface is still smaller than that
of the 1-2 interface, according to Fig.~\ref{f:sigma}(a) (the line 
$g_{12} = 1.1$), and therefore, component 3 flows further into the $x > x_0$
region as time elapses.
Figure~\ref{f:marangoni}(c) shows the case where $g_{12}$ is changed from
1.01 to 1.02 at $t = 0$.
In this case, no Marangoni-like flow is observed, even though there is a
gradient in interfacial tension.
This is because, for $g_{12} = 1.02$, the interfacial tension increases with
$n_3$, as shown in Fig.~\ref{f:sigma}(a);
even if the 1-3-2 interface were formed, its interfacial tension would be
larger than that of the 1-2 interface, and the 1-3-2 interface would be
pulled back in the $-x$ direction.
Thus, the flow of component 3 into the 1-2 interface is suppressed.

\section{Conclusions}
\label{s:conc}

We have investigated the static and dynamic properties of an interface in an
immiscible three-component BEC.
We considered the case illustrated in Fig.~\ref{f:schematic}, where
component 3 was sandwiched by components 1 and 2.
We calculated the interfacial tension of the three-component interface and
showed that a component 3 inserted into the 1-2 interface can lower the
interfacial tension, i.e., it can play the role of surfactant
(Fig.~\ref{f:sigma}).
We obtained the dependence of the interfacial tension on various
parameters (Figs.~\ref{f:sigma} and \ref{f:delta}).
We proposed a variational wave function for understanding the surfactant
behavior qualitatively (Sec.~\ref{s:var}).
The stability of the three-component interface was examined by Bogoliubov
analysis (Fig.~\ref{f:bogo}), and it was shown that the three-component
interface is dynamically stable when component 3 plays the role of
surfactant.
Finally, we studied the case shown in Fig.~\ref{f:marangoni} to demonstrate
that an interfacial tension gradient induces a Marangoni-like flow.
We showed that an interface with smaller interfacial tension is entrained
toward an interface with larger interfacial tension, resulting in 
interfacial flow.

An extension of this study would be to focus on emulsification in turbulent
superfluids.
When oil and water are stirred, the large interfacial tension between oil
and water prevents their droplets from becoming small.
By adding a surfactant, the interfacial tension between oil and water
decreases, allowing the droplets to become smaller, i.e., emulsified.
We would expect to see a similar trend in a stirred three-component BEC upon
the addition of the third component.
This phenomenon may be realized given the recent developments in
quantum-turbulence experiments in BECs~\cite{Navon1,Navon2}.

\begin{acknowledgments}
This work was supported by JSPS KAKENHI Grant Number JP20K03804.
\end{acknowledgments}

\appendix

\section{Calculation of the variational grand potential}
\label{app}

We calculate the grand potential in Eq.~(\ref{Omega}) using the variational
wave function in Eq.~(\ref{var}).
Substituting Eq.~(\ref{var}) into the kinetic energy,
\begin{equation}
  E_j^{\rm (kin)} = -\frac{1}{2} \int_{-\infty}^\infty dz \psi_j^{\rm (var)*}
  \frac{d^2\psi_j^{\rm (var)}}{dz^2},
\end{equation}
we obtain
\begin{equation} \label{a1}
  E_1^{\rm (kin)} = E_2^{\rm (kin)} = \frac{1}{8 \alpha}
\end{equation}
and
\begin{equation} \label{a2}
  E_3^{\rm (kin)} = \frac{1}{4\alpha^2} \left[ \alpha \coth \frac{w}{\alpha}
  - w \left( \sinh \frac{w}{\alpha} \right)^{-2} \right].
\end{equation}
Since we are considering an infinite space, the intracomponent interaction
and the chemical potential part,
\begin{equation} \label{Eintra}
  E_j^{\rm (intra)} = \int_{-\infty}^\infty dz \left(
  \frac{g_{jj}}{2} |\psi_j^{\rm (var)}|^4
  - \mu_j |\psi_j^{\rm (var)}|^2 \right),
\end{equation}
diverge for $j = 1$ and 2, and we must subtract the divergent part from
them.
For $j = 1$, the indefinite integral of Eq.~(\ref{Eintra}) is calculated to
be (we are assuming $g_{11} = 1$ and $\mu_1 = 1$)
\begin{eqnarray}
& & \int dz \left( \frac{1}{2} |\psi_j^{\rm (var)}|^4
- |\psi_j^{\rm (var)}|^2 \right) \nonumber \\
& = & -\frac{z}{4} + \frac{\alpha}{4} \ln \cosh \frac{z + z_0}{\alpha}
- \frac{\alpha}{8} \tanh \frac{z + z_0}{\alpha},
\end{eqnarray}
which approaches
\begin{subequations}
\begin{eqnarray}
  \label{div}
  & & -\frac{z}{2} + \frac{1}{8}[\alpha (1 - \ln 4) - 2 z_0]
  \qquad (z \rightarrow -\infty), \\
  & & \frac{1}{8} [\alpha (-1 - \ln 4) + 2 z_0]
  \qquad (z \rightarrow \infty).
\end{eqnarray}
\end{subequations}
The divergence stems from the term $-z / 2$ in Eq.~(\ref{div}).
Since this divergent part does not include the variational parameters, we can
drop this term in the variational analysis.
Thus, the regularized form of the definite integral in Eq.~(\ref{Eintra})
becomes
\begin{equation} \label{a3}
E_1^{\rm (intra)} = \frac{1}{4}(2 z_0 - \alpha).
\end{equation}
Since $\psi_2^{\rm (var)}(z) = \psi_1^{\rm (var)}(-z)$, we have the same
result for $j = 2$: $E_2^{\rm (intra)} = (2 z_0 - \alpha) / 4$.
For $j = 3$, there is no divergence, and we obtain
\begin{equation} \label{a4}
  E_3^{\rm (intra)} = \frac{1}{2} \left( w \coth \frac{w}{\alpha} - \alpha
  \right) - \mu_3 w.
\end{equation}
The integrals $I_{jj'} \equiv \int_{-\infty}^\infty dz
|\psi_j^{\rm (var)}|^2 |\psi_{j'}^{\rm (var)}|^2$ in the intercomponent
interaction energies are calculated to be
\begin{equation} \label{a5}
  I_{12} = \frac{2z_0}{e^{4 z_0 / \alpha} - 1},
\end{equation}
and
\begin{equation} \label{a6}
  I_{13} = I_{23} = \frac{1}{2} \left(
  \frac{w - 2 x_0}{1 - e^{(2 z_0 - w) / \alpha}} 
  + \frac{w + 2 x_0}{1 - e^{(2 z_0 + w) / \alpha}} \right).
\end{equation}
The variational grand potential $\Omega_{\rm var}$ is thus obtained from
Eqs.~(\ref{a1}), (\ref{a2}), (\ref{a3}), (\ref{a4}), (\ref{a5}), and
(\ref{a6}):
\begin{equation}
  \Omega_{\rm var} = \sum_{j=1}^3 \left( E_j^{\rm (kin)} + E_j^{\rm (intra)}
  \right) + g_{12} I_{12} + g_{13} I_{13} + g_{23} I_{23}.
\end{equation}


\begin{thebibliography}{99}
\bibitem{Note}
We are not including spinor BECs in this discussion.
For the interfaces of spinor BECs, see for example, Refs.~\cite{Borgh,
  Borgh2, Kaneda}.
\end{thebibliography}
\end{document}